\documentclass[aps,pra,twocolumn,groupedaddress,showpacs]{revtex4}

\newcommand{\myfigurewidth}{\columnwidth}
\newcommand{\calion}[1]{$^{#1}$Ca$^+$}
\newcommand{\calatom}[1]{$^{#1}$Ca}

\newcommand{\degcelcius}{$^\circ\!\!\;$C}
\newcommand{\msqdiff}[1]{$\delta\langle r^2\rangle ^{#1}$}

\usepackage{latexsym} \usepackage{amssymb}\usepackage{amsmath}
 \usepackage{epsfig}

\begin{document}


\title{Isotope shifts of the $4s^2$ $^1S_0 \to 4s5p$ $^1P_1$
  transition and\\
 hyperfine splitting of the $ 4s5p$ $^1P_1$ state
  in calcium}


\author{A.~Mortensen}\email[Email address: ]{andersvm@phys.au.dk}
\author{J.~J.~T.~Lindballe} \author{I.~S.~Jensen} \author{P.~Staanum}
\author{D.~Voigt}\altaffiliation{Present address: Huygens Laboratory, University
  of Leiden, The Netherlands.}
 \author{M.~Drewsen}

\affiliation{QUANTOP, Danish National Research Foundation Center for
  Quantum Optics, \\
Department of Physics and Astronomy, University of Aarhus, DK-8000
  Aarhus C, Denmark}


\date{\today}

\begin{abstract}
  
  Using a technique based on production of ion Coulomb crystals,
  the isotope shifts of the $4s^2$ $^1S_0\to4s5p$ $^1P_1$ transition
  for all naturally occurring isotopes of calcium as well as the
  hyperfine splitting of the $4s5p$ $^1P_1$ state in \calatom{43} have
  been measured. The field shift and specific mass shift coefficients
  as well as the hyperfine structure constants for \calatom{43} have
  been derived from the data.
\end{abstract}
\pacs{42.62.Fi, 32.10.Fn, 32.80.Pj}


\maketitle

\section{Introduction}

Isotope shifts and hyperfine splitting of optical transitions provide
valuable information about atomic electron configurations and not at
least properties of the nuclei. While from isotope shifts, nuclear
charge distributions can be deduced (see,
e.g., Refs~\cite{Heilig:1974,Nadjakov:1994}), from hyperfine splittings,
nuclear spins, magnetic dipole moments, and electric quadrupole moments
can be determined~\cite{Casimir:1936}.

Although isotope shifts and hyperfine splittings already have been
measured for a large number of transitions in neutral \calatom{} (see Refs.~\cite{Palmer:1984,Nortershauser:1998,Mueller:2000} and references
therein) and in singly charged
\calion{} ions~\cite{Maleki:1992,Kurth:1995,Alt:1997}, the $4s^2$
$^1S_0\to4s5p$ $^1P_1$ transition considered in the present paper
has hitherto not been measured. In contrast to previous experiments,
the present measurements use ion Coulomb crystals as means to obtain
the spectroscopic data.

\section{Experimental procedure}
\begin{figure}[htb]
  \centering
  \epsfig{figure=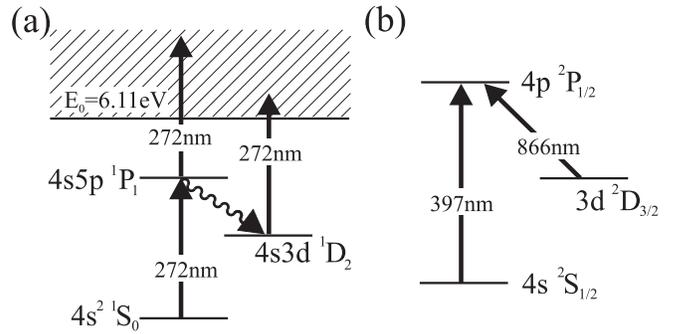,width=\myfigurewidth}
   \caption{Simplified energy level diagrams of Ca (a) and \calion{}
     (b) of relevance to the experiments.  The straight arrows
     indicate laser driving transitions, while the wavy arrow
     symbolizes spontaneous emission. In (a), the solid line beneath
     the hatched region indicates the ionization limit.}
\label{fig:levelscheme}

\end{figure}

\begin{figure}[htb]
  \centering
  \epsfig{figure=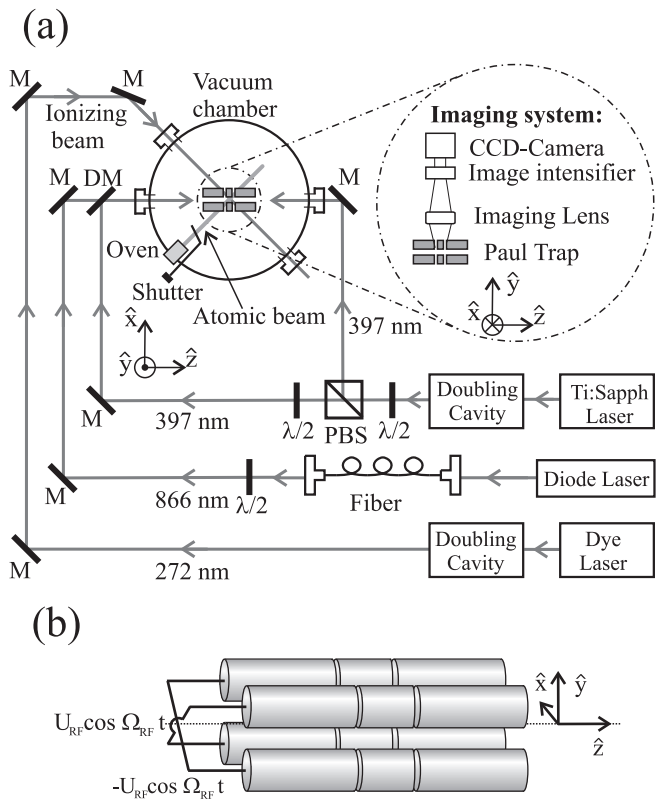,width=\myfigurewidth}
   \caption{(a) Schematics of the experimental setup with an inset
  showing a side view of the
     imaging system. For the optical elements in the figure, the
     following notation has been used: M, mirror; DM, dichroic mirror;
     PBS, polarization beam splitter; and $\lambda
     /2$, half-wave-plate. (b) The linear Paul trap electrodes with
     applied rf voltages. The $z$ axis
    defines the trap axis. The middle sections are 5.4 mm long and the
    end pieces are 20 mm each. The diameter of the electrodes is 8.0
    mm and the minimum distance to the trap axis is 3.5 mm.}\label{fig:setup}
\end{figure}

In the experiments, we have used a frequency doubled cw laser system
tuned to the $4s^2$ $^1S_0\to4s5p$ $^1P_1$ transition at a wavelength
of $\lambda \approx 272$ nm. The degree of excitation at a particular
wavelength is determined from the ion production rate when excited
neutral \calatom{} atoms absorb yet another $\lambda \approx 272$ nm
photon directly from the $4s5p$ $^1P_1$ state or from the metastable
$4s3d$ $^1D_0$ state, populated through spontaneous emission as
depicted in Fig.~\ref{fig:levelscheme}(a)~\cite{Smith:1988}. As shown
in the sketch of the experimental setup in Fig.~\ref{fig:setup}(a),
the spectroscopic laser beam at $\lambda \approx 272$ nm crosses a
well-collimated effusive thermal beam of calcium atoms at right angles
in the center of a linear Paul trap~\cite{Prestage:1989}. This
geometry is chosen in order to obtain small Doppler shifts of the
transition under study as well as to have a large capture efficiency
of the ions produced. Since the atomic beam is derived from an oven
containing natural metallic calcium, all the isotopes \calatom{40}
(96.9\%), \calatom{42} (0.647\%), \calatom{43} (0.135\%), \calatom{44}
(2.086\%), \calatom{46} (0.004\%), and \calatom{48} (0.187\%) are
present. The detection of the ions produced within a given exposure
period is done by first trapping them independently of the specific
isotope, and then converting almost all ions into \calion{40} through
near-resonant electron transfer collisions with atoms in the atomic
beam containing 96.9\% \calatom{40} atoms.  Finally, the number of
trapped ions is counted with a near-100\% efficiency by monitoring the
fluorescence from the \calion{40} ions when laser cooled into a
Coulomb crystal~\cite{Diedrich:1987,Wineland:1987}. Below, the
key parts of the applied technique are discussed in detail.

\subsection{Isotope and hyperfine selective ion production}
 
The laser system used for the spectroscopy consists of a tunable,
single-mode ring CW dye laser operated at 544 nm with an output power
of a few hundreds mW which is frequency doubled to 272 nm by a
$\beta$-Barium Borate crystal placed in an external cavity. The output power at
the desired $4s^2$ $^1S_0\to 4s5p$ $^1P_1$ transition wavelength is
typically about $\sim$ 10 mW, and the laser linewidth is below 1 MHz.
This linewidth is much smaller than the full width at half maximum(FWHM) Doppler width of $\sim$
50 MHz due to the transverse velocity spread of the thermal atomic
beam originating from an oven at a temperature of $\sim
600$\degcelcius. Hence, since the natural linewidth of the $4s^2$
$^1S_0\to4s5p$ $^1P_1$ transition according to the NIST
database~\cite{NISTdata} is known to be $\Gamma\approx 2 $ MHz
($>$ 50\% uncertainty), the Doppler effect is the resolution limiting
factor in the experiments. With isotope shifts typically of the order
of GHz and hyperfine splittings of about $\sim$ 100 MHz of
\calatom{43}, not only can one selectively excite individual even
isotopes (no nuclear spin), but also the three hyperfine components of
the $4s5p$ $ ^1P_1$ state in the case of \calatom{43}. As mentioned
above, ions are produced from excited calcium atoms either by
absorbing yet another $\lambda \approx 272$ nm photon directly from
the $4s5p$ $^1P_1$ state, or from the metastable $4s3d$ $^1D_0$ state
populated through spontaneous decay. By exposing the atomic beam to
the photoionization laser beam with a certain frequency for a fixed
period of time (in our experiment controlled by a mechanical light
shutter), the ion production will reflect the excitation degree of the
$4s5p$ $^1P_1$ state of calcium atoms at this frequency. Counting the
number of produced ions as a function of laser frequency thus reveals
the excitation spectrum.

\subsection{The ion detection scheme}

The central element of the ion detection is the linear Paul trap
situated in a vacuum chamber which is operated at a pressure of $\sim 10^{-10}$
Torr. As sketched in Fig.~\ref{fig:setup}(b), the trap consists of
four sectioned cylindrical electrodes placed in a quadrupole
configuration. Confinement of ions in the $xy$ plane is obtained by an
effective radial harmonical potential created by applying a
rf-potential of amplitude $U_{RF}$ and frequency $\Omega_{RF}$ to
each of the electrodes in such a way that the voltage on the
diagonally opposite electrode is in phase while the voltage of the
neighbouring electrodes is 180$^\circ$ out of phase. Along the trap
axis, the $z$-axis, confinement originates from a positive dc voltage
$U_{EC}$ applied to all the eight end-pieces. With the trap dimensions
used [see Fig.~\ref{fig:setup}(b)], for $U_{RF} \sim 200$ V, $\Omega =
2 \pi \times 3.9$ MHz and $U_{EC} \sim 5$ V, the effective trap depth
becomes $\sim 1$ eV. This ensures that ions produced in the trapping
region are efficiently trapped despite the thermal velocities of the
neutral atoms.

\begin{figure}[htb]
  \centering
  \epsfig{figure=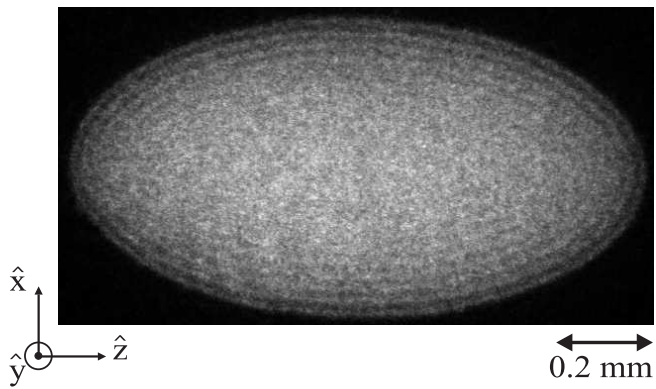,width=\myfigurewidth}
  \caption{Image of a pure \calion{40} Coulomb crystal consisting of
  $\sim 10\,000$ ions.}
  \label{fig:cleancryst}
\end{figure}

When ions of a certain calcium isotope, e.g., \calion{40}, are
produced, they can be Doppler laser-cooled by driving the $4s$
$S_{1/2}\to 4p$ $P_{1/2}$ and the $3d$ $D_{3/2}\to 4p$ $P_{1/2}$
transitions by laser light around 397 nm and 866 nm, respectively, as
depicted in Fig.~\ref{fig:levelscheme}(b). As the ions are cooled to
sufficiently low temperatures ($\sim 10$ mK), they arrange in a
spatial ordered structure, often referred to as an ion Coulomb
crystal. In our setup, these ion Coulomb crystals are imaged onto an
image intensified Charge-coupled device camera by collecting fluorescence from the
Doppler cooled ions. A typical picture of a Coulomb crystal containing
$\sim 10\,000$ \calion{40} ions is shown in Fig.~\ref{fig:cleancryst}.
Here, one clearly observes the elliptical projection of the crystal,
which for our trapping potential is spheroidal with the trap axis
($z$ axis) as the axis of rotational symmetry.  Single-component
Coulomb crystals in our trap have spatially uniform ion densities when
viewed on length scales larger than the typical distance between
neighboring ions (see Ref.~\cite{Hornekaer:2001}). The specific density is set by the
trapping parameters. Hence, by determining the crystal volumes,
e.g.,~by measuring the main axis of the projected ellipse, we can
determine the total number of trapped ions.

\begin{figure}[htb]
  \centering \epsfig{figure=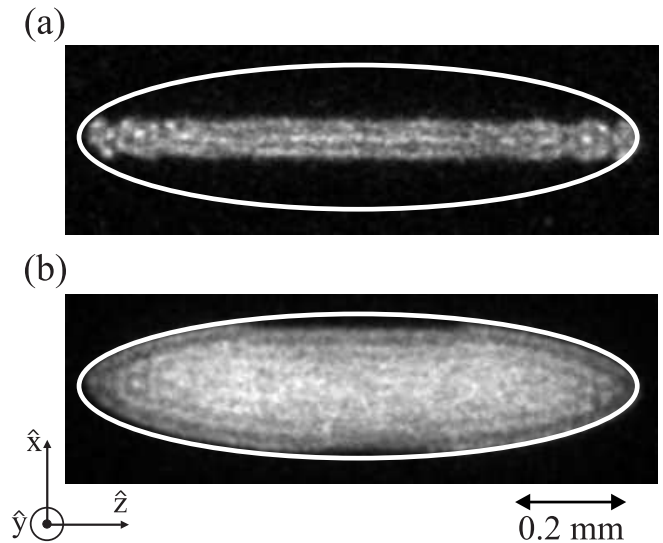,width=\myfigurewidth}
  \caption{Image of a
    \calion{40}(fluorescing)--\calion{44}(nonfluorescing)
    two-component crystal of $\sim 1\,500$ ions before (a)
    and after (b) exposure to the atomic beam which leads to the
    electron transfer as described in the text. The white line
    indicates the outer boundary of the crystal.}
  \label{fig:electronseq}
\end{figure}

Generally, when the frequency of the spectroscopy laser is tuned
between the resonances the various isotopes, two or more singly
charged calcium isotope ions will be produced and trapped. Such
multispecies ion ensembles can be cooled into Coulomb
crystals as well, with the lighter isotopes more tightly bound towards the
trap axis than the heavier due to the dynamical confinement in the
$xy$ plane~\cite{Hornekaer:2001}.  In Fig.~\ref{fig:electronseq}(a), an
image of the fluorescence from laser cooled \calion{40} ions in such a
crystal is shown. The crystal consists mainly of \calion{40}- and
\calion{44} ions, where the nonfluorescing \calion{44} ions are
cooled only sympathetically through the Coulomb interaction with the
directly laser cooled \calion{40} ions \cite{Bowe:1999,Molhave:2000}.
The number of ions in the multicomponent crystal could principally be
determined by imaging the fluorescence from the individual isotopes.
This is, however, technically very demanding, since each isotope ion
would need its own set of laser frequencies due to isotope shifts of
the transitions shown in Fig.~\ref{fig:levelscheme}(b). As an
alternative route to quantify the total ion production, we choose to
expose the multicomponent crystals similar to the one in
Fig.~\ref{fig:electronseq}(a) to the thermal atomic beam of calcium
until nearly all the ions ($\sim$ 96.9\%) by near-resonant electron
transfer collisions of the type
\begin{equation}
^A\textrm{Ca}^+  + ^{40}\textrm{Ca} \to ^A\textrm{Ca} + ^{40}\textrm{Ca}^+
\label{transfer}
\end{equation}
have been converted into \calion{40} ions, and subsequently we measure
the size of the now nearly pure \calion{40} crystal. Since the energy
difference between the two sides of Eq.~(\ref{transfer}) is just the
isotope shift of the electronic levels, which is much smaller than the
thermal energy of the atoms in the beam, no significant energy barrier
exists for the process. Furthermore, since the kinetic energy of the
produced \calion{40} ions is more than an order of magnitude smaller
than the trap potential depth, essentially 100\% of the ions produced
through Eq.~(\ref{transfer}) stay trapped. In
Fig.~\ref{fig:electronseq}(b), the two-component crystal presented in
Fig.~\ref{fig:electronseq}(a) has been converted via the electron
transfer process into a nearly pure \calion{40} crystal.

\section{Spectroscopic results}

\subsection{The $4s5p$ $^1P_1$ excitation spectrum}

\begin{figure}[htb]
  \centering \epsfig{figure=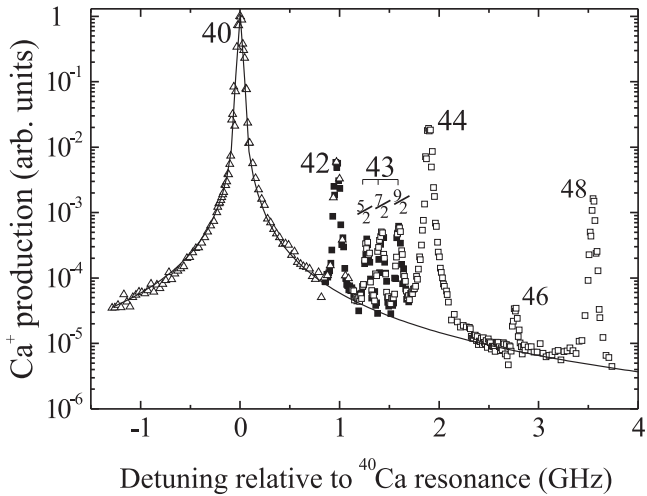,width=\myfigurewidth}
  \caption{Semilogarithmic plot of the production rate of
    \calion{} ions as a function of the frequency detuning of the
    spectroscopy laser from the \calatom{40} resonance. The data
    points presented are the results of three frequency scans made at
    slightly different temperatures [$(\vartriangle)$ $T_{oven} =
    612^\circ$C, $(\blacksquare)$ $T_{oven} = 630^\circ$C, 
    and $(\square)$ $T_{oven} = 602^\circ$C]. The full line indicates
    a Voigt profile fit to the \calatom{40} resonance within the
    frequency range $\sim [-1300;700]$ MHz.}
  \label{fig:scangraph}
\end{figure}

Figure~\ref{fig:scangraph} shows the relative ion production rate or,
equivalently, the relative rate at which calcium atoms are being
excited to the $^1P_1$ state, as a function of the detuning of the
spectroscopy laser. These relative rates were obtained from knowledge
of the volumes of the created Coulomb crystals, the laser exposure
times, and the intensities of the spectroscopy laser. The spectrum
contains resonances of all the naturally abundant isotopes, including
the one corresponding to \calatom{46} which has a natural abundance
of only 0.004\%.  The results presented in Fig.~\ref{fig:scangraph}
have been obtained from three partially overlapping frequency scans.
The three data series were measured at slightly different oven
temperatures: $T_{oven} = 612^\circ$C $(\vartriangle)$, $T_{oven} =
630^\circ$C $(\blacksquare)$, and $T_{oven} = 602^\circ$C $(\square)$.
The relative rates of the $\vartriangle$ data and the
$\blacksquare$ data have been normalized, so they share the same
fitted maximum value and position at the \calatom{42} resonance peak.
In the same way the data indicated by $\blacksquare$ and $\square$
have been normalized to the three \calatom{43} hyperfine peaks.

The temperature was during each experiment kept stable to better than
$\pm 2^\circ$C corresponding to a maximum uncertainty in the atomic density
of about $\pm 7\%$ during the whole data accumulation time of $\sim40$
min. Since the time to measure a single resonance is only a very small
fraction of this time, any systematic errors in the resonance profiles
due to fluctuating oven temperatures can be neglected. The Doppler
broadening of the resonances, which only depends slightly on the
atomic mass, is best found by fitting the \calatom{40} resonance data
to a Voigt profile. By doing this one obtains a total FWHM of
$\sim 54\pm 5$ MHz dominated by the Doppler broadening. At the laser
intensity which we apply ($\sim 100$ mW/cm$^2$), we have assumed that
neither the $4s^2$ $^1S_0\to4s5p$ $^1P_1$ transition nor the following
photoionization process is saturated. Consequently, the ion
production rate is expected to be proportional to the laser intensity
squared. We accounted for fluctuations in the spectroscopic laser
intensity during the data accumulation time by using this assumption.
That this assumption is rather good is supported by noting that the
relative measured resonance peak values in Fig.~\ref{fig:scangraph}
are in good agreement with the natural abundances of the calcium
isotopes, which are \calatom{40} (96.9\%), \calatom{42} (0.647\%),
\calatom{43} (0.135\%), \calatom{44} (2.086\%), \calatom{46} (0.004\%),
and \calatom{48} (0.187\%).

\subsection{Isotope shifts}\label{isotopeshifts}

Before deriving the isotope shifts from data like the ones presented
in Fig.~\ref{fig:scangraph}, the uncertainty of the individual data
points has to be estimated and taken into account. The uncertainty is
essentially due to three effects. First, due to intensity fluctuations
of the laser intensity there will be a fluctuation in the number of
ions produced. Since for neighboring data points around a resonance
this intensity fluctuation is maximally a few percent, the expected
square dependence on the intensity of the ion production leads to an
estimated uncertainty of $\sim 5 \% $. Second, in determination of the
Coulomb crystal volumes, both systematic and random uncertainties due
to the measurements of the main axis of the elliptical projections of
the crystals occur. Both these errors are of about $\sim 5 \% $, but
since the systematic errors are equal for points symmetrically
positioned around a resonance, it will have a small effect on the
determination of the resonance frequencies. Hence, in our analysis, we
have only accounted for the random errors of $5 \% $. Finally, due to
the finite number of ions produced within one measurement, we have
included an uncertainty of the square root of the estimated number of
ions. If all these uncertainties are accounted for in a weighted least-squares Gaussian fit, the uncertainty on the resonance position
becomes lower than 2 MHz.

A more critical error arises when we compare similar scans. Here we
find that the measured resonance peak positions are associated with
much larger uncertainties than the ones from the Gaussian fits. This
additional error originates from local frequency drift of the
spectroscopy laser during a whole scan which typically lasts about 40
min. From a series of scans, we have found that this laser drift error
leads to a rms uncertainty in the resonance frequencies of about 9 MHz.
This last error source is the dominant one for the estimated errors of
the isotope shifts given in Table~\ref{tab:isotopeshifts}. In this
table, the isotope shift for \calatom{43} is found as the center of
gravity of the three hyperfine components of \calatom{43} which will
be discussed in Sec.~\ref{hyperfine}.  In addition to these
statistical errors, there is also a systematic uncertainty of $\pm
1\%$ arising from the calibration of the spectroscopy laser frequency
scan to an optical spectrum analyzer with a known free-spectral range.

For completeness, we have estimated that we might have introduced an
unimportant error of a few hundred kHz due to the fact that we have
determined the resonance frequencies from single Gaussian fits instead
as from a more realistic multipeak Voigt profile fit.

\begin{table}[htb]
  \centering
  \begin{tabular}{@{\hspace{1em}}c@{\hspace{2em}}c@{\hspace{1em}}} 
    \hline
    \hline
    Mass  $(A)$& Shift (MHz)\\
    \hline   
    42&  $967 \pm 9$\\
    43& $1455 \pm 9$\\
    44& $1879 \pm14 $\\
    46& $2746\pm16 $\\
    48& $3528 \pm16$\\
    \hline
    \hline
  \end{tabular}
  \caption{Isotope shifts of the transition $4s^2$ $^1S_0 \to 4s5p$ $^1P_1$ in calcium derived from the experimental data like the ones presented in Fig.~\ref{fig:scangraph}. All shifts are with respect to the \calatom{40} resonance. The errors stated represent one standard deviation originating from the statistical errors in the experiments. In addition, the
     shifts are subject to an overall linear scaling uncertainty of $1\%$ due to our frequency scan calibration (see
     text). The isotope shift for \calatom{43} is the center of gravity of the hyperfine components.}
   \label{tab:isotopeshifts}
 \end{table}

The isotope shift for a given transition is usually described as a sum
of the mass and the field shift in the following
way~\cite{Heilig:1974}
\begin{equation} 
\delta \nu^{AA'}=M\frac{A'-A}{AA'}+ F\delta \langle r^2\rangle^{AA'},
\end{equation}
where $M$ is the mass shift coefficient, $A$ and $A'$ denote the
atomic masses of the two isotopes, $F$ is the field shift coefficient,
and \msqdiff{AA'} is the difference in mean square nuclear charge
radii between the isotopes.

\begin{figure}[htb]
  \centering
  \epsfig{figure=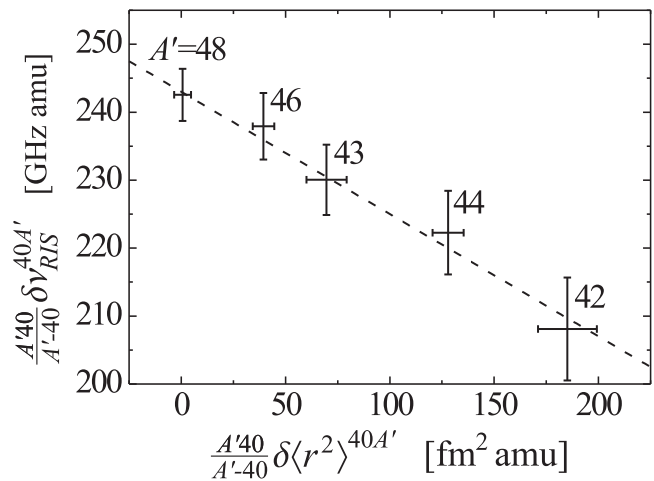,width=\myfigurewidth}
  \caption{
    Plot of the values $[A'40/(A'-40)]\delta\nu^{40A'}_{RIS}$
    derived from the measurements as function of
    $[A'40/(A'-40)]$\msqdiff{40A'} for the $4s^2$ $^1S_0 \to 4s5p$
    $^1P_1$ transition in \calatom{}. The dashed line is a weighted
    linear regression fit to determine the coefficients $M_{SMS}$ and
    $F$ from the relation given in Eq.~(\ref{eq:1}).  }
  \label{fig:mRIS}
\end{figure}

The mass shift is usually written as a sum of the normal mass shift
(NMS) and the specific mass shift (SMS), which means that we can write
the mass shift coefficient as $M = M_{NMS} + M_{SMS}$. Here the NMS
coefficient is given by the simple expression $M_{NMS}= \nu_0
m_e/m_u$, where $\nu_0$ is the transition frequency, $m_e$ is the
electron mass, and $m_u$ is the atomic mass unit. The NMS originates
from the reduced mass correction for the electron, while the SMS comes
from the change in the correlated motion of all the electrons (see,
e.g., Ref.~\cite{Sobelman:1972}).  Subtraction of the NMS from the total
isotope-shift gives the residual isotope shift (RIS),
\begin{equation}
\label{eq:5}
\delta\nu^{AA'}_{RIS} = M_{SMS} (A'-A)/AA' +  F\delta \langle r^2\rangle^{AA'}.
\end{equation}
Rewriting Eq.~(\ref{eq:5}) by multiplication by the factor
$AA'/(A'-A)$ leads to
 \begin{equation}
   \label{eq:1}
  \frac{A'A}{A'-A} \delta\nu^{AA'}_{RIS} = M_{SMS} +  F  \left(\frac{ A'A}{A'-A}\delta \langle r^2\rangle^{AA'}\right),
 \end{equation}
 which shows that $M_{SMS}$ and $F$ can be determined from a linear
 fit when the \msqdiff{AA'}'s are known. Using values of the
 \msqdiff{AA'}'s for calcium from Ref.~\cite{Nadjakov:1994}, in
 Fig.~\ref{fig:mRIS}, we have plotted the left-hand side of
 Eq.~(\ref{eq:1}) using the values of Table~\ref{tab:constants} as a
 function of $[A'A/(A'-A)]$\msqdiff{AA'} for the fixed value of
 $A=40$. The SMS coefficient and field shift coefficient, obtained by
 weighted linear regression fit to the data points in
 Fig.~\ref{fig:mRIS}, are listed in Table~\ref{tab:constants}.

\begin{table}[htb]
  \centering
  \begin{tabular}{@{\hspace{1em}}c@{\hspace{2em}}c@{\hspace{1em}}} 
    \hline
    \hline
    $M_{SMS}$
    (GHz amu)&
    $F$ (MHz/fm$^2$)
    \\
    \hline
  
    $243\pm 3  \pm 9 $&           
    $-179\pm 39 \pm 2$           
    \\
    \hline
    \hline
  \end{tabular}
 \caption{The specific mass shift $M_{SMS}$ 
and field shift $F$ coefficients for the $4s^2$ $^1S_0
    \to 4s5p$ $^1P_1$ transitions of calcium derived from the the linear fit presented in Fig.~\ref{fig:mRIS}. The first stated uncertainty estimate is the one standard deviation obtained
    from the linear regression to the data of Fig.~\ref{fig:mRIS}. The
    systematic error in the isotope shifts due to the frequency scan
    calibration is included as the second uncertainty estimate. For
    completeness the NMS coefficient is $M_{NMS}=604.3$ GHz amu. }
  \label{tab:constants}
\end{table}

The field shift coefficient for the $4s^2$ $ ^1S_0 \to 4s5p$ $^1P_1$
transition (Table~\ref{tab:constants}) is, within the stated error,
almost equal to the experimentally determined field shift coefficient
for the $4s^2$ $^1S_0 \to 4s4p$ $^1P_1$ transition of $F = -175.8 \pm
1.2$ MHz/fm$^2$  reported in Ref.~\cite{Nortershauser:1998}. This can be
attributed to the fact that a $p$ electron has negligible overlap with
the nucleus compared to an $s$ electron, so the main contribution to
the field shifts is expected in both cases to come from the
$s$ electrons.

\subsection{\calatom{43} hyperfine splitting}\label{hyperfine}

The nuclear spin of $I=7/2$ for \calatom{43} leads to three hyperfine
levels of the $^1P_1$ state with total spins $F= 5/2, 7/2$, and $9/2$,
respectively.  The hyperfine structure (hfs) constants and isotope
shift of \calatom{43} are determined by fitting to the normal Casimir
formula~\cite{Casimir:1936}
\begin{equation}
  \label{eq:2}
\begin{split}  
  \Delta E_F=&
     \Delta \nu _{cg} + \frac{A}{2} C\\
    & + \frac{B}{4}
    \frac{\frac{3}{2}C(C+1)-2I(I+1)J(J+1)}{(2I-1)(2J-1)IJ} ,
\end{split}
\end{equation}
where $C= F(F+1) -I(I+1) -J(J+1)$, $\Delta\nu_{cg}$ is the
isotope shift of the center of gravity of the hfs, and $A$ and $B$ are
the magnetic dipole and electric quadrupole coupling constants,
respectively.  Several scans across the three hyperfine resonances
were made to increase the level of confidence of the $A$ and $B$
constants. The hfs constants derived from these scans are summarized
in Table~\ref{tab:HFS}, while the center of gravity for \calatom{43}
has been given in Table~\ref{tab:isotopeshifts}.  The small value of
the $B$ constant indicates that the magnetic dipole coupling has the
most prominent contribution to lifting the degeneracy of the $4s5p$
$^1P_1$ level for \calatom{43}. Compared with the work of
Ref.~\cite{Nortershauser:1998}, where the hfs constants for the $4s4p$
$^1P_1$ state in \calatom{43} have been measured to be $A= -15.54 \pm
0.03$ MHz and $B = -3.48 \pm 0.13$ MHz, the $4s5p$ $^1P_1$ state has
an opposite sign for the $A$ constant and the $B$ constant is the same
order of magnitude or smaller.

\begin{table}[htb]
  \centering
  \begin{tabular}{@{\hspace{1em}}c@{\hspace{2em}}c@{\hspace{1em}}} 
    \hline
    \hline
    $A$ (MHz)&
    $B$ (MHz)\\
    \hline
    $39.8 \pm 0.8 \pm 0.4$&
    $-0.3 \pm 3  \pm 0.03$ 
    \\
    \hline
    \hline
  \end{tabular}
  \caption{hfs constants for the $4s5p$ $^1P_1$ state in \calatom{43}. The first stated uncertainties
    originate from statistical errors in determining the resonance positions, while the second account for the
    systematic errors due to frequency calibration uncertainty.}
  \label{tab:HFS}
\end{table}

\section{Conclusion}

With a technique based on determining the number of ions produced,
collected, and cooled into Coulomb crystals through resonant two-photon
ionization, the isotope shifts of the $4s^2$ $^1S_0\to4s5p$ $^1P_1$
transition for all naturally occurring isotopes of calcium as well as
the hyperfine splitting of the $4s5p$ $^1P_1$ state in \calatom{43}
have been measured. The field shift and specific mass shift
coefficients as well as the hyperfine structure constants for
\calatom{43} have been derived from the data. Though in the presented
experimental scheme, near-resonant charge exchange collisions were
used to gain information about the ion number, it is not a requirement
for the technique for these to be used. Actually, the technique should
be applicable to many atomic species as long as some laser cooled ions
are simultaneously trapped and can sympathetically cool the species of
interest into a Coulomb
crystal~\cite{Larson:1986,Molhave:2000,Drewsen:2003}. From the spatial
organization of the observable laser cooled species, one can easily
deduce the number of the, e.g., nonfluorescing ions of interest.

\section{Acknowledgements}
This work was supported by the Danish National Research Foundation and
the Carlsberg Foundation.

\bibliography{mybib}

\end{document}